\documentclass[aps,prl,twocolumn,showpacs,superscriptaddress,amsmath,amssymb]{revtex4-1}

\usepackage{xcolor}
\usepackage{graphicx}
\usepackage{multirow}
\usepackage{booktabs}
\usepackage{colortbl}
\usepackage{hhline}
\usepackage[left,modulo]{lineno}
\usepackage{upgreek}
\usepackage[
			colorlinks=true,
			urlcolor=blue,
			linkcolor=blue,
			citecolor=blue,
			filecolor=blue
			pagebackref=blue,
			]{hyperref}

\newcommand\etal{\textit{et al}}

\begin{document}

\title{Bright luminescence from indirect and strongly bound excitons in hBN}

\author{L\'eonard Schu\'e}
\affiliation{Laboratoire d'Etude des Microstructures, ONERA-CNRS, Universit\'e Paris-Saclay, BP 72, 92322 Ch\^atillon Cedex, France}
\affiliation{Groupe d'Etude de la Mati\`ere Condens\'ee, UVSQ-CNRS, Universit\'e Paris-Saclay, 45 avenue des Etats-Unis, 78035 Versailles Cedex, France}
\author{Lorenzo Sponza}
\affiliation{Laboratoire d'Etude des Microstructures, ONERA-CNRS, Universit\'e Paris-Saclay, BP 72, 92322 Ch\^atillon Cedex, France}
\author{Alexandre Plaud}
\affiliation{Laboratoire d'Etude des Microstructures, ONERA-CNRS, Universit\'e Paris-Saclay, BP 72, 92322 Ch\^atillon Cedex, France}
\affiliation{Groupe d'Etude de la Mati\`ere Condens\'ee, UVSQ-CNRS, Universit\'e Paris-Saclay, 45 avenue des Etats-Unis, 78035 Versailles Cedex, France}
\author{Hakima Bensalah}
\affiliation{Groupe d'Etude de la Mati\`ere Condens\'ee, UVSQ-CNRS,  Universit\'e Paris-Saclay, 45 avenue des Etats-Unis, 78035 Versailles Cedex, France}
\author{Kenji Watanabe}
\affiliation{National Institute for Materials Science, 1-1 Namiki, Tsukuba 305-0044, Japan}
\author{Takashi Taniguchi}
\affiliation{National Institute for Materials Science, 1-1 Namiki, Tsukuba 305-0044, Japan}
\author{Fran\c{c}ois Ducastelle}
\affiliation{Laboratoire d'Etude des Microstructures, ONERA-CNRS, Universit\'e Paris-Saclay, BP 72, 92322 Ch\^atillon Cedex, France}
\author{Annick Loiseau}
\affiliation{Laboratoire d'Etude des Microstructures, ONERA-CNRS, Universit\'e Paris-Saclay, BP 72, 92322 Ch\^atillon Cedex, France}
\author{Julien Barjon}
\affiliation{Groupe d'Etude de la Mati\`ere Condens\'ee, UVSQ-CNRS,  Universit\'e Paris-Saclay, 45 avenue des Etats-Unis, 78035 Versailles Cedex, France}
\email{julien.barjon@uvsq.fr, annick.loiseau@onera.fr}


\begin{abstract}
A quantitative analysis of the excitonic luminescence efficiency in hexagonal boron nitride (hBN) is carried out by cathodoluminescence in the ultraviolet range and compared with zinc oxide and diamond single crystals. A high quantum yield value of $\sim$50\% is found for hBN at 10 K comparable to that of direct bandgap semiconductors. This bright luminescence at 215 nm remains stable up to room temperature, evidencing the strongly bound character of excitons in bulk hBN. \textit{Ab initio} calculations of the exciton dispersion confirm the indirect nature of the lowest-energy exciton whose binding energy is found equal to 300$\pm$50 meV, in agreement with the thermal stability observed in luminescence. The direct exciton is found at a higher energy but very close to the indirect one, which solves the long debated Stokes shift in bulk hBN.
\end{abstract}

\maketitle

Hexagonal boron nitride (hBN) is a wide bandgap semiconductor ($>$6 eV) that has recently encountered a tremendous regaining of interest with the emergence of the two-dimensional (2D) crystals family \cite{Novoselov2004,Xia2014}. Because of its lattice isostructural to graphene and its excellent dielectric properties, it has been designated as the best insulating material for improving electron mobility in graphene \cite{Dean2010} or enhancing intrinsic optical properties of transition metal dichalcogenides (TMDs) \cite{Cadiz2017,Ajayi2017}.

In the past decade, the growth of high quality hBN single crystals has also opened new possibilities for light-emitting devices. Pioneering works by Watanabe \etal. have revealed the high UV radiative efficiency of hBN at room temperature \cite{Watanabe2004} and have led to the design of the first hBN-based light emission device in the deep ultraviolet range \cite{Watanabe2009a}. However, the physical mechanisms behind hBN luminescence are still debated. The main controversy revolves around two apparently contradictory aspects: on the one hand, the highly efficient luminescence (up to 10$^3$ times higher than in diamond at 300 K) and on the other hand, the energy shift between absorption and emission (Stokes shift).

Among the various interpretations proposed in the past, Watanabe \etal. first came to the conclusion that hBN luminescence was driven by direct excitonic recombinations \cite{Watanabe2004}, and they attributed the Stokes shift to the lifted degeneracy of free exciton levels due to a dynamical lattice distortion \cite{Watanabe2009}. But recently, Cassabois \etal. attributed the luminescence lines to phonon-assisted transitions from an indirect exciton \cite{Cassabois2016}. The interpretation is based on a weakly bound hydrogenic picture of the exciton, the Wannier-Mott model, whose limits have already been emphasized for excitons in TMD atomic layers \cite{Qiu2013,Chernikov2014,Wang2018}. It also does not look pertinent in hBN where \textit{ab initio} calculations predict a direct exciton close to the Frenkel type \cite{Wirtz2005,Arnaud2006}.

Until now, although being widely established, the high luminescence efficiency (LE) of hBN has never been quantitatively investigated. The classical methods using photoluminescence (PL) combined with integrating spheres \cite{Crosby1971,Jacquez1955,Ware1965} are not adapted for the deep UV range as required for wide bandgap semiconductors such as hBN.

In this Letter, we report on a quantitative study of the cathodoluminescence (CL) in hBN compared to two wide bandgap semiconductors, zinc oxide and diamond. A careful calibration of our detection setup gives access to the absolute CL intensities in the 200-400 nm range. For the first time, low-temperature measurements (10 K), as a function of excitation depth, provide a value of the internal quantum yield (QY) in a bulk hBN single crystal. Temperature-dependent CL from 10 K up to 300 K further show the high stability of excitons in bulk hBN compared to diamond, and this attests of their strong binding energy. These findings are found consistent with \textit {ab initio} calculations performed in this work to establish the dispersion of excitons in hBN. They have encouraged us to revise the fundamental optical properties of excitons in hBN, marked by a long debated Stokes shift. A consistent picture describing exciton luminescence and absorption energies is proposed.\\

For hBN, all measurements were performed on a single crystal provided by the NIMS (Japan), grown at high pressure high temperature (HPHT) \cite{Taniguchi2007}, the highest quality source available today. The diamond and ZnO crystals, as well as the experimental CL set-up, are described in Supplemental Material I. The intensity calibration of the CL detection system relies on a deuterium lamp with a calibrated spectral irradiance (LOT Oriel 30 W) used as a measurement standard within the 200-400 nm wavelength range. Once corrected for the spectral response of the setup, absolute CL intensities are obtained. The light emission power is assessed through a careful evaluation of optical, geometrical, and sample parameters. The full LE measurement procedure including the absorbed power determination is detailed in Supplemental Material I. Note that, the uncertainty related to the absolute LE measurements in the deep UV by CL ($\pm $ 50\%) is higher than when using integrating spheres in the visible range. 

\begin{figure}[h!]
 \begin{center}
 \includegraphics[scale=0.85]{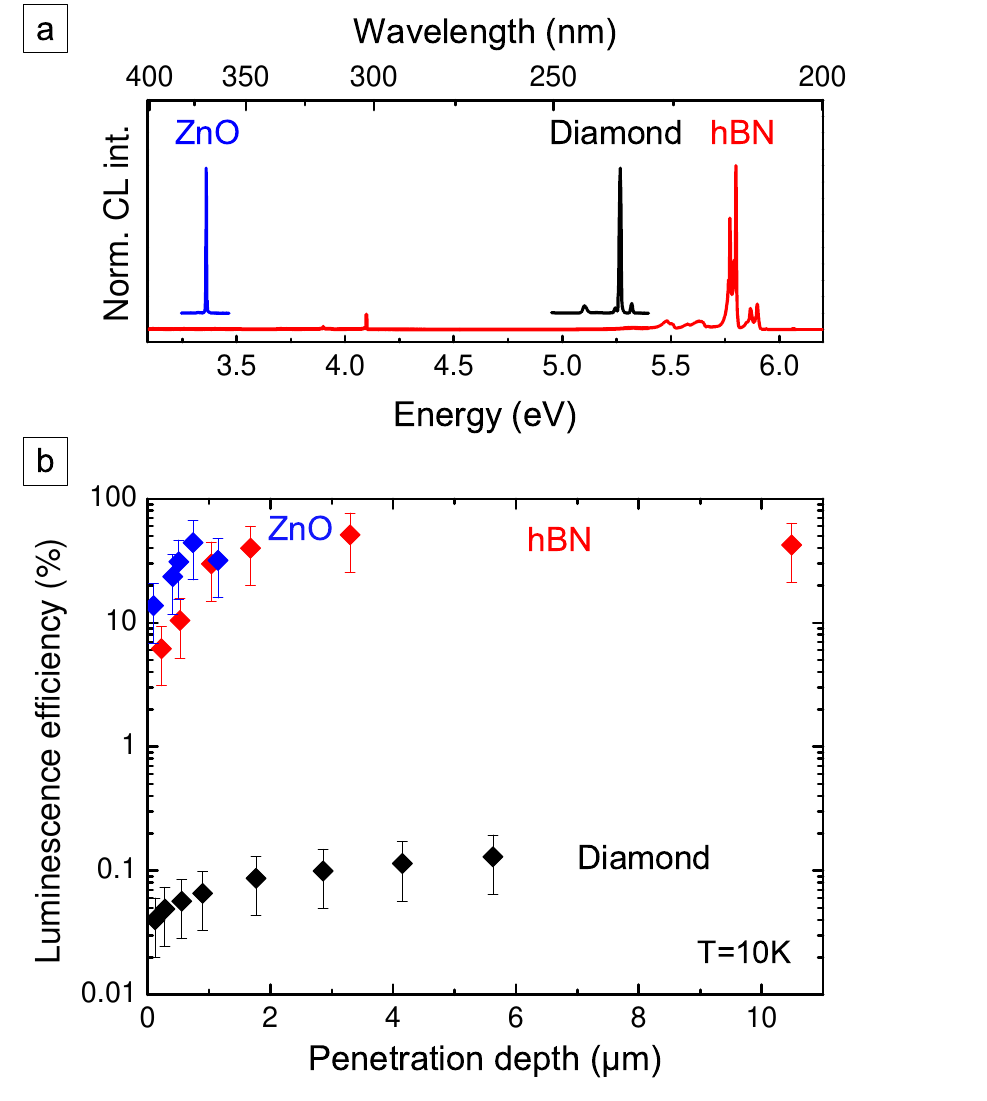}
 \caption{a) Cathodoluminescence spectra of hBN, diamond and ZnO recorded in the 200-400 nm range at 10 K. For clarity, the excitonic spectra of ZnO and diamond are upshifted. All spectra are corrected from the spectral response of the detection system. b) Luminescence efficiencies of ZnO, hBN and diamond plotted as a function of electron penetration depth. Accelerating voltages from 3 to 30 kV. T=10 K. The incident power $Vi$ is kept constant to avoid non-linear effects: 1 $\upmu$W, 6 $\upmu$W and 0.4 $\upmu$W for hBN, diamond, and ZnO respectively.}
 \label{F1}
 \end{center}
\end{figure}

Prior to any quantitative analysis, an overview of the low temperature CL signal is given in the UV range (200-400 nm) for hBN, diamond, and ZnO single crystals (Fig.\ref{F1}a). The luminescence of ZnO is dominated by a sharp peak at 3.37 eV, corresponding to the recombinations of neutral-donor bound excitons \cite{Klingshirn2010}. In contrast, the luminescence occurs from free excitons in diamond and hBN. In diamond, it presents a series of lines detected around 5.25 eV coming from phonon-assisted (TA, TO, LO) recombinations of indirect free excitons \cite{Dean1965,Barjon2017}. Similarly, the UV spectrum of hBN displays a series of sharp lines with a maximum at 5.795 eV, recently attributed to recombinations of indirect free excitons \cite{Cassabois2016}. Note that, in both samples, the luminescence from deep defects is almost undetectable allowing a proper analysis of the intrinsic excitonic features.

The luminescence efficiencies were measured at increasing accelerating voltages (3-30 kV). In CL, high-energy incident electrons undergo a successive series of elastic and inelastic scattering events in the crystal before being completely stopped. The penetration depth of electrons $R_{p}$ then increases with the electron beam energy $V$ following the empirical relationship of Kanaya \etal. \cite{Kanaya1972}, $R_{p}$ ($\upmu$m) = $C.V$(kV)$^{1.67}$ where $C$ is a material-dependent parameter.

Figure \ref{F1}b presents the luminescence efficiencies plotted as a function of the penetration depth. While at high $R_{p}$ the LE saturates in all cases, one observes a significant luminescence quenching at low excitation depths. In $sp^3$ semiconductors such as ZnO or diamond, dangling bonds ending crystal surfaces are known to introduce non-radiative recombination channels that lower the radiative efficiency. In 2D crystals with $sp^2$ hybridization dangling bonds are not expected, but the surface effects still remain poorly known. Possible explanations relying on contaminated or defective surface \cite{Amani2015} or Auger effects \cite{Mouri2014} can be invoked to account for this result.

At high penetration depths, the luminescence efficiency reaches a constant value called the internal quantum yield (QY), which is characteristic of the bulk crystal. Generally noted $\upeta$, it corresponds to the fraction of excitons that recombines radiatively inside the crystal. While the LE most often depends on the crystal orientation, surface terminations and contaminations, the internal quantum yield remains a reference parameter of the bulk material.

As expected, the direct bandgap ZnO crystal is found to be highly radiative with a QY larger than 50\%, in good agreement with previous PL analysis \cite{Hauser2008}. Conversely, the internal quantum yield of diamond remains close to 0.1\%, more than two orders of magnitude lower than in ZnO. Indirect excitons formed in diamond require the simultaneous emission of a phonon to recombine radiatively, which inevitably lowers the probability of such a process. In the case of hBN, the LE also saturates at high penetration depths but it reaches an $\sim$50\% internal quantum yield, comparable to that of a direct semiconductor such as ZnO. Such a high QY value from phonon-assisted recombinations of indirect excitons far exceeds what is commonly observed for indirect semiconductors. It indicates that the non-radiative channels present in the crystal are efficiently by-passed by faster radiative recombinations (a few hundred picoseconds according to the literature \cite{Watanabe2011b}).

\begin{figure}[h!]
 \begin{center}
 \includegraphics[scale=0.81]{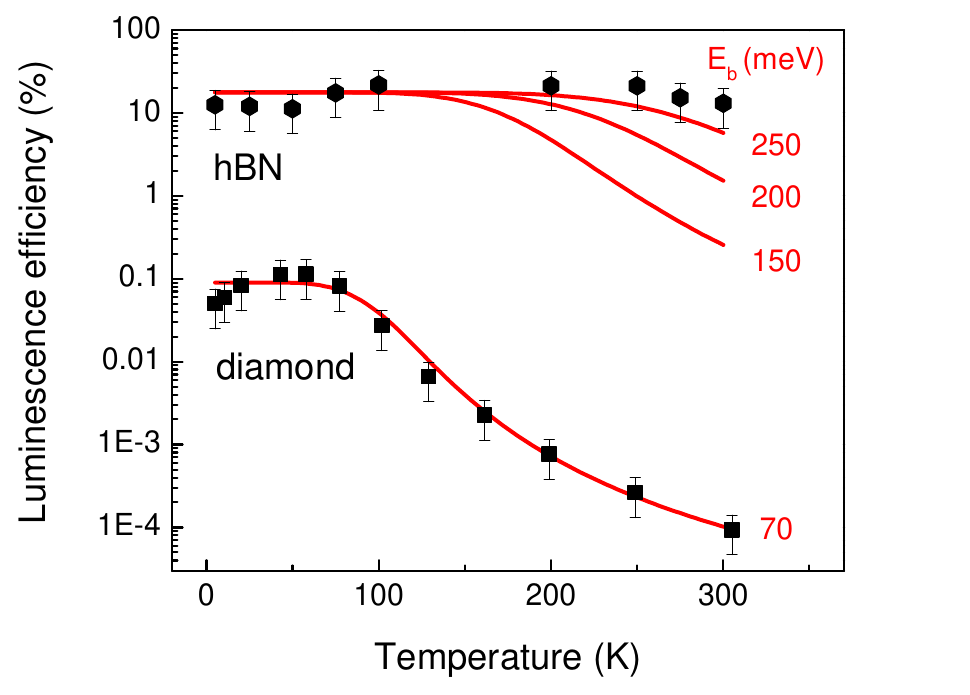}
 \caption{Luminescence efficiency (logarithmic scale) of hBN and diamond as a function of temperature. Excitation parameters: $V$=10 kV, $i$=0.2 nA for hBN and 9.5 nA for diamond. The curves (red lines) correspond to the thermal dissociation of excitons (See Supplemental Material II) where $E_{b}$ is the exciton binding energy.}
 \label{F2}
 \end{center}
\end{figure}

We then performed temperature-dependence measurements of the luminescence efficiency. Figure \ref{F2} first presents the CL results obtained on the two indirect bandgap semiconductors, namely hBN and diamond from 10 K to 300 K. To limit surface effects, the accelerating voltage was set at 10 kV. In diamond, the QY is rather stable at low temperatures because of the predominance of the exciton population over free electrons and holes. Above 60 K the thermal dissociation of excitons occurs and the quasi-equilibrium between exciton and free carrier populations is displaced in favor of free carriers resulting in a drastic LE drop of 3 orders of magnitude. This temperature behavior is well described by the classical model for exciton dissociation in silicon \cite{Gourley1982,Labrie1985} recently implemented in diamond \cite{Kozak2014}, with $E_{b}$ = 70 meV in agreement with the reference values of the literature \cite{Clark1964,Dean1964}.

The situation is much different in hBN where the LE remains almost constant over the full temperature range. At 300 K, the exciton population is still predominate, demonstrating that the exciton binding energy in hBN is much higher than in diamond. By applying the same thermal dissociation model (see Supplemental Material II), the binding energy of the lowest-energy excitons in bulk hBN is estimated to be larger than 250 meV. This lower bound is much higher than the previously reported values in the literature (149 meV \cite{Watanabe2004} and 128 meV \cite{Cassabois2016}).\\

\begin{figure}[t]
 \begin{center}
 \includegraphics[scale=0.81]{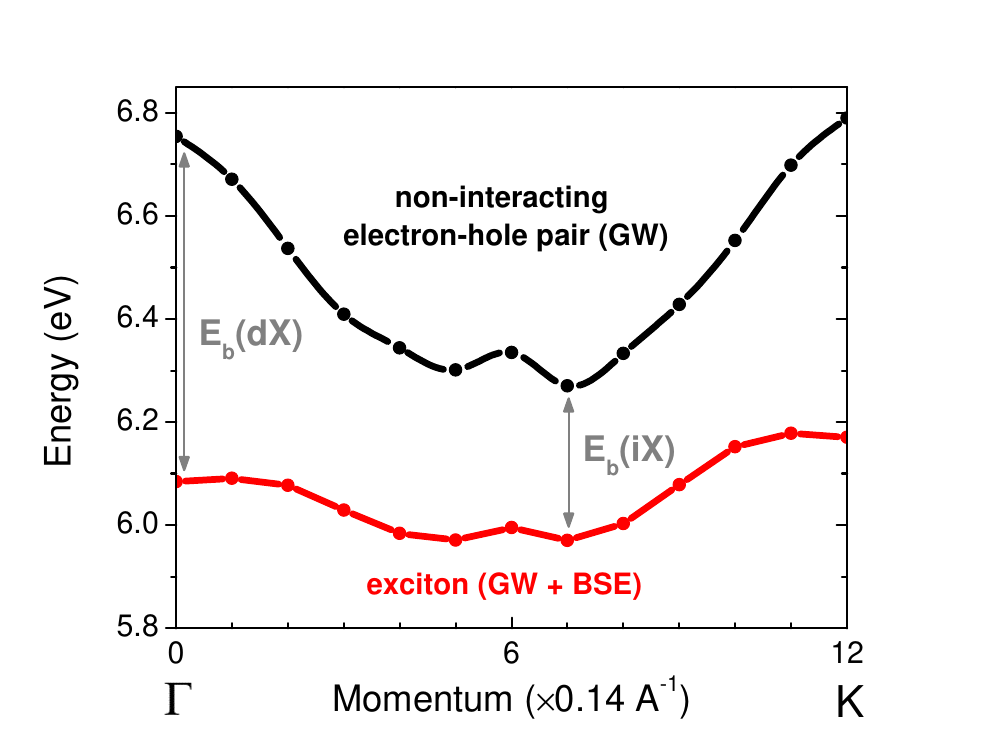}
 \caption{Dispersion of excitons in hBN compared to the non-interacting electron-hole pair along the $\Gamma$K direction. The vertical arrows indicate the binding energy $E_b$ for the direct ($dX$) and indirect ($iX$) excitons. Adapted from Ref. \cite{Sponza2018}.}
 \label{F3}
 \end{center}
\end{figure}

To elucidate this controversy, the \textit{ab initio} Bethe-Salpeter equation (BSE) has been solved to obtain the exciton dispersion E$_{exc}(Q)$, with $Q$ being the exciton momentum. Quasiparticle energies have been computed within the perturbative GW method and further blue shifted by 0.47 eV on the basis of recent electron energy loss spectroscopy (EELS) experiments \cite{Schuster2018}. The state-dependent corrections of GW ensure an accurate dispersion at the single-particle level. Details of the computational method are given in Supplemental Material III. Two different two-particle dispersion relations can then be obtained. First, the dispersion of the non-interacting electron-hole pair (GW, black curve) has been traced by reporting the lowest energy difference between the empty and occupied quasiparticle states compatible with the given $Q = k_e + k_h$. Second, the exciton dispersion (GW-BSE, red curve) results from the solution of the BSE at finite $Q$. Electron-hole pairs formed with electrons sitting in M at the bottom of the conduction band and holes sitting near K at the top of the valence band are expected to have a momentum close to KM = 1/2 $\Gamma$K \cite{Cassabois2016}. That is why we report in Figure \ref{F3} both non-interacting e-h pair and exciton dispersions along the $\Gamma$K direction.

The independent e-h dispersion (GW curve) reaches its lowest energy near the middle of $\Gamma$K and lies 0.5 eV below the direct transition ($\Gamma$), consistently with the indirect electron band structure (see Supplemental Material III and Ref.\cite{Arnaud2006}). In the presence of Coulombic interactions, the minimum of the exciton energy (GW-BSE curve) is also found close to the middle of $\Gamma$K at 5.97 eV. The exciton dispersion behavior corroborates the attribution of the luminescence lines as being phonon-assisted recombinations of an indirect exciton \cite{Cassabois2016}. It is also confirmed through high-resolution CL spectroscopy presented in Supplemental Material I. The value of 5.956 eV is found by CL for the no phonon recombination, which accurately defines the indirect exciton ($iX$) energy in hBN.

More insights come from the comparison of the exciton and free electron-hole pair dispersions (Fig.\ref{F3}). Strikingly, excitonic effects tend to flatten the dispersion of the non-interacting electron-hole pair. In other terms, the exciton is found to be much heavier than the sum of the electron and hole masses, a sum that is assumed to be the exciton mass in the Wannier-Mott model. As a first consequence, the binding energies of direct and indirect excitons are different. The binding energy of the indirect exciton ($E_b(iX)$) is found equal to 300$\pm$50 meV (at $Q = \frac{7}{12} \Gamma$K), consistent with the high thermal stability observed experimentally (Fig.\ref{F2}). Incidentally, from the energies found for $E_b(iX)$ and $iX$, the indirect bandgap energy in bulk hBN is refined to $E_g=$ 6.25$\pm0.05$ eV. Instead, at $Q=0$, the direct exciton binding energy $E_b(dX)$ is found to be equal to $670$ meV, close to the previously reported values \cite{Wirtz2005,Arnaud2006}). The second consequence of the exciton dispersion flattening is that the direct and indirect excitons are very close in energy. The direct exciton lies only $\sim$100 meV above the indirect one.

\begin{figure}[h!]
 \begin{center}
 \includegraphics[scale=0.96]{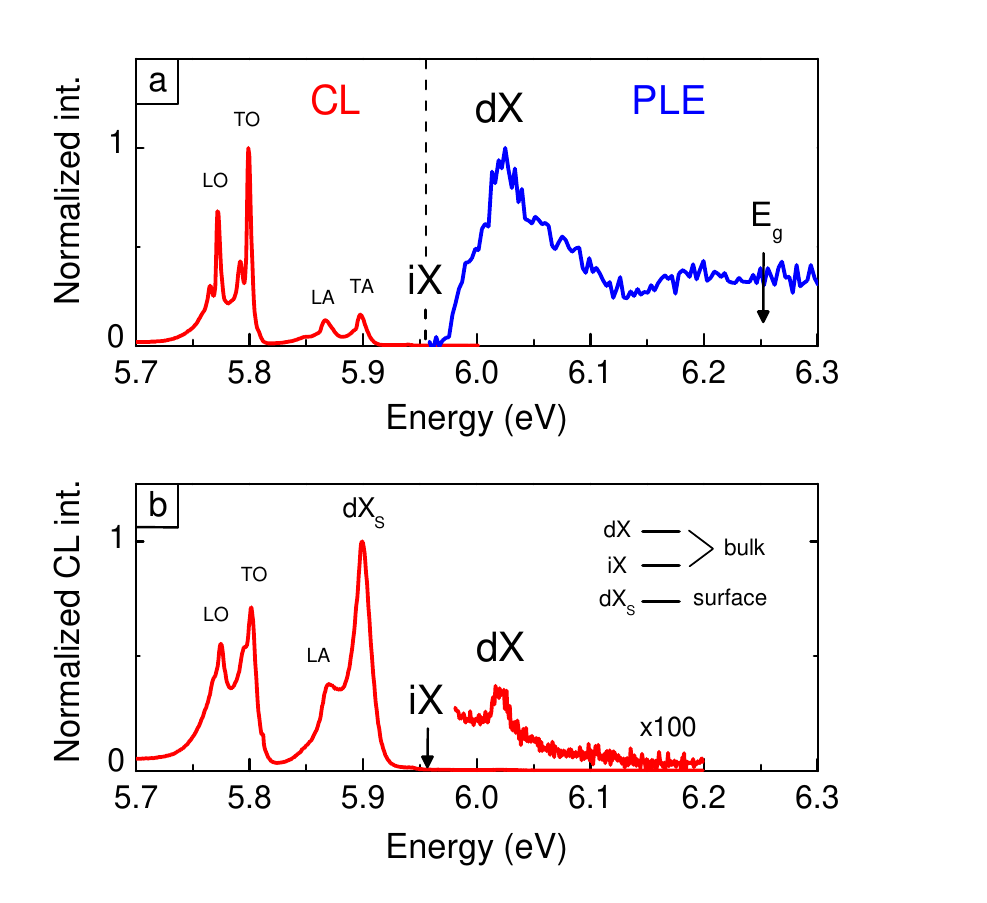}
 \caption{a) Normalized spectra of CL and PLE recorded at the CL maximum (from Ref. [\citenum{Museur2011}]) illustrating the Stokes shift observed in bulk hBN crystals. b) CL spectrum of a thin exfoliated hBN flake showing the radiative recombination of non-thermalized direct excitons detected at 6.03 eV. CL and PLE spectra taken at 10 K.}
 \label{F4}
 \end{center}
\end{figure}

This theoretical work sheds a new light on the origin of the Stokes shift reported for hBN between the maxima of luminescence and absorption energies \cite{Watanabe2004,Museur2011} as illustrated in Figure \ref{F4}a with the CL and photoluminescence excitation (PLE) (from Ref. \cite{Museur2011}) spectra recorded on the same crystal. It has been reported that the sharp PLE signal peaking at 6.025 eV would arise from a phonon-assisted process \cite{Cassabois2016}. In indirect semiconductors, one should expect a mirror symmetry between absorption and luminescence around the $iX$ energy position \cite{Elliott1957}, both in energy and intensity. It is, however, not observed in hBN, and it does not apply to interpreting the absorption spectrum.

On the other hand, Figure \ref{F4}a shows that the energy difference between the $iX$ and the PLE maximum energies, about 70 meV, is comparable to the theoretical $\sim$100 meV energy difference between direct and indirect excitons (Fig.\ref{F3}). This lets us propose that the dominant PLE peak arises from the direct exciton labeled $dX$ in Figure \ref{F4}a. Al$_x$Ga$_{1-x}$As alloys face a similar situation when $x \sim 0.3$ but with weakly-bound excitons ($E_b \sim$ 5 meV) \cite{Monemar1976}. The luminescence of hBN which only reveals the lowest energy states is driven by indirect exciton ($iX$) recombinations while in the PLE spectrum, being similar to an absorption one, the direct exciton ($dX$) is predominate.

Even though high resolution absorption data are still lacking, all available experiments exhibit a main peak around 6.1 eV with a strong absorption coefficient, higher than $10^{5}$ cm$^{-1}$ \cite{Zunger1976,Stenzel1996,Vuong2017mbe}. Such a behavior is typical of optical transitions without phonons (in silicon, for instance, $\alpha$ is higher than $10^{5}$ cm$^{-1}$ at the direct gap edge \cite{Philipp1960}). Furthermore, recent studies have shown that the absorption spectrum of hBN multilayers is fairly independent of the number of layers \cite{Kim2013,Liu2013}. The direct exciton is theoretically described as being quite similar in bulk hBN and in the monolayer, with its energy being weakly influenced by the surrounding BN planes \cite{Wirtz2006,Wirtz2009,Galvani2016}.

Finally, CL experiments done on exfoliated hBN samples have also provided precious information. Figure \ref{F4}b displays a typical CL spectrum recorded on an 80 nm thick hBN flake, where the bulk contribution of indirect excitons (TO, LO, LA, and ZA phonon-assisted lines according to the assignments proposed in Ref. \cite{Cassabois2016}) is detected together with a sharp peak at 5.90 eV. In a previous work \cite{Schue2016}, we have already shown that when reducing the thickness, a progressive vanishing of the bulk luminescence lines is accompanied by the emergence of a narrow single emission at this energy. This single emission peak tends to indicate that the associated luminescence process does not involve phonons and originates from a direct optical transition. It is further attributed here to surface excitons ($dX_{S}$) with zero momentum and an energy slightly below the indirect exciton of bulk hBN, according to recent theoretical works \cite{Paleari2018}. More interestingly for the present discussion, in Figure \ref{F4}b a weak luminescence peak could be detected at 6.03 eV, the same energy as the PLE maximum. It is attributed to the recombination of non-thermalized (hot) direct excitons ($dX$) from bulk hBN, observed simultaneously with the TO, LO, LA replica of the indirect one ($iX$). Finally, the co-existence of direct and indirect excitons with close energies is confirmed as a key to understanding the luminescence and absorption properties of hBN.

In summary, the luminescence quantum yield in bulk hBN is quantitatively estimated to be $\sim$50\% as high as in direct bandgap semiconductors such as zinc oxide. \textit{Ab initio} calculations of the excitonic dispersion have been decisive in providing a consistent picture of exciton properties in bulk hBN. The lowest-energy excitons involved in luminescence are of an indirect nature, with a 300$\pm$50 meV binding energy, consistent with the stability of the luminescence intensity observed up to room temperature. The excitonic dispersion has revealed the presence of a direct exciton at a slightly higher energy, responsible for the maximum of absorption in bulk hBN. The long-standing debate on the Stokes Shift in bulk hBN is thus completely elucidated. As a perspective, the extremely high efficiency of the phonon-assisted luminescence in hBN, unique for an indirect bandgap semiconductor, still needs a better understanding for applications with light emitting devices in the deep UV range.\\

\acknowledgments

Christ\`{e}le Vilar is warmly acknowledged for the technical help on the cathodoluminescence-SEM setup. Authors would like to thank the French National Agency for Research (ANR) for funding this work under the project GoBN (Graphene on Boron Nitride Technology), Grant No. ANR-14-CE08-0018. The research leading to these results has also received funding from the European Union Seventh Framework Program under grant agreements no. 696656 GrapheneCore1 and no.785219 GrapheneCore2. Growth of hexagonal boron nitride crystals was supported by the Elemental Strategy Initiative conducted by the MEXT, Japan and JSPS KAKENHI Grant Numbers JP15K21722.

\bibliography{Bibliography}

\end{document}


\title{\Large{Supplemental Material}}

\maketitle

\section{Section I.}
 \subsection{A. Samples and experimental setup}
 
The cathodoluminescence setup combines an optical detection system (Horiba Jobin Yvon SA) and a JEOL7001F field-emission gun scanning electron microscope (FEG-SEM) equipped with a Faraday cup for the beam current measurement. Samples are mounted on a GATAN cryostat SEM-stage to cool them at temperatures from 300 K down to 5 K thanks to cold-finger cryostat with liquid helium. The temperature reported in this work is measured on the sample holder. The CL signal is collected by a parabolic mirror and focused with mirror optics on the open entrance slit of a 55 cm-focal length monochromator. The all-mirror optical setup provides an achromatic focusing with a high spectral sensitivity down to 190 nm. A nitrogen-cooled charge-coupled detector (CCD) silicon camera is used to record CL spectra. The spectral calibration is performed using a Hg lamp.\\

The diamond and ZnO single crystals used in our experiments are of high quality and commercially available: Element6 (Electronic Grade CVD, impurities $<$10$^{14}$ at/cm$^{3}$, dislocation density $<$10$^{5}$ /cm$^{2}$) and CrysTec Gmbh (impurities $\sim$10$^{17}$ at/cm$^{3}$) respectively. Note that the hBN crystal is less pure, with carbon and oxygen impurities detected by secondary ion mass spectroscopy in the 10$^{18}$ at/cm$^{3}$ range (see Ref. [18] in the main text). The luminescence from deep defects near 4.1 eV remains several orders of magnitude weaker than exciton ones. A 200-1000 nm spectrum of the same sample is presented in our previous work [1].

\begin{figure}[h!]
\renewcommand\thefigure{SI}
 \begin{center}
 \includegraphics[scale=0.95]{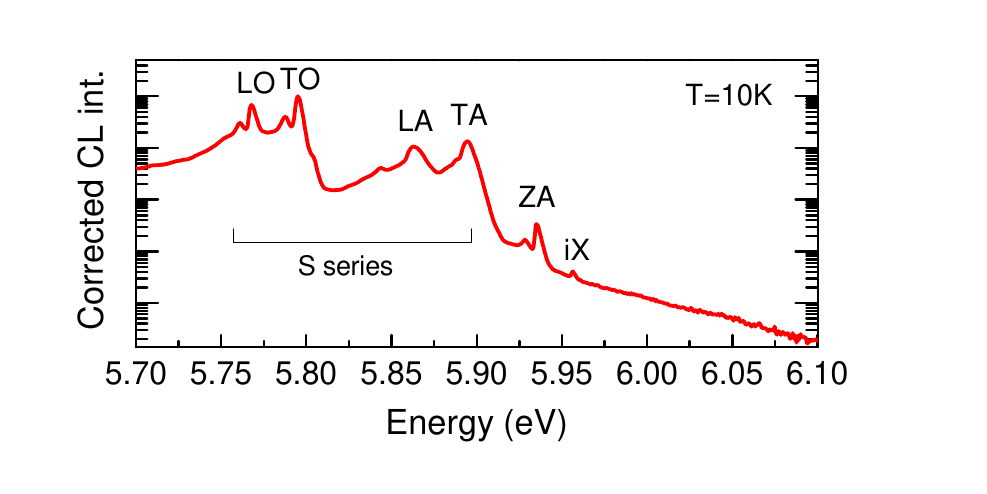}
 \caption{High resolution CL spectrum of a HPHT-grown hBN crystal recorded in the near band edge region (logarithmic scale).}
 \label{FS1}
 \end{center}
\end{figure}

Figure \ref{FS1} displays a high-resolution CL spectrum of the hBN single crystal at 10 K, showing the same emission energies as those obtained with two-photon PL excitation [2]. The series of peaks at 5.768(S4), 5.795(S3), 5.864(S2), 5.894(S1) and 5.935 eV initially reported as the S-series corresponds to phonon-assisted recombinations of the indirect exciton as proposed in Ref. [2]. Importantly, we observe that the radiative recombination without phonon (labeled $iX$) occurs at 5.956 eV which, accurately defines the indirect exciton energy in hBN.

 \subsection{B. Measuring the luminescence efficiency by cathodoluminescence}

One of the fundamental differences between CL and PL is that, whereas one absorbed photon generates one coherent electron hole (e-h) pair in PL, one single electron generates hundreds of incoherent e-h pairs in CL. The generation rate G (\textit{i.e.}, the number of e-h pairs generated in the crystal per unit of time) is given by:

\begin{equation}
	\centering
		G = \frac{{V}{i}(1-f)}{<E>e}
		\label{eq-G}
\end{equation}

\noindent
where $V$ is the accelerating voltage of incident electrons, $i$ the beam current, $e$ the electronic charge and $f$ the backscattering factor. $<$\textit{E}$>$ is the average energy required for the formation of an electron–hole pair in the crystal. Here, we assume that $<$\textit{E}$>$ is equal to 3$E_{g}$ [3]. This empirical relation is known to account for the electron-hole formation energy with a typical uncertainty of 25\% (see for instance the measured value of $<$\textit{E}$>$ in diamond in Ref. [4]). Given the further uncertainties on $V$ and $i$ measured with a Faraday cup, the generation rate is evaluated at $\pm$30 \%. 

Under continuous excitation, the steady state population of excitons inside the crystal is equal to $G\tau$, with $\tau$ the exciton lifetime. The total photon flux emitted inside the crystal is equal to $G\tau$/$\tau_{\rm rad}$, $\tau_{\rm rad}$ being the radiative lifetime of excitons. The ratio of the light emission over the generation rate defines here the luminescence efficiency LE = $\tau$/$\tau_{\rm rad}$. 

However, in a luminescence experiment, photons are collected outside the crystal so that corrections have to be made to determine the luminescence efficiency ($\tau$/$\tau_{\rm rad}$) from absolute CL intensities. One has to consider how a photon escapes the crystal (light extraction efficiency) as well as instrumental parameters. The flux of photons collected by the detector from the specimen is given by the following equation assuming an isotropic light emission [5]:

\begin{equation}
    \centering
    I_{\rm{CL}} = \frac{G\tau}{\tau_{\rm{rad}}} . (\Omega/4\pi)F_{A}F_{R}F_{I}
    \label{eq:I-CL}
\end{equation}

\noindent
Where:

\begin{itemize} 

\item $\Omega$ is the solid angle of collection of the detection system limited by the parabolic mirror aperture. In this work, it is equal to 3.9 steradians which corresponds to a $\Omega$/4$\pi$ = 18\% collection of the emitted light outside the crystal.
\item $F_{R}$ is the light extraction factor which corrects internal reflections and refraction at the specimen surface. $F_{R}$ is an analytic function of the refractive index $n$ [6]. We took $n$(hBN) = 3.95 at 215 nm [7], $n$(diamond) = 2.67 at 236 nm [8] and $n$(ZnO) = 2.26 at 372 nm [8]. 
\item $F_{A}$ accounts for the luminescence re-absorption along the escape path of photons. It is neglected in semiconductors exhibiting large Stoke shifts between light absorption and emission, such as diamond and hBN. Though re-absorption is significant for a direct bandgap semiconductors such as ZnO, it was also neglected in this work. Note this assumption results in underestimating the luminescence efficiency of ZnO.
\item $F_{I}$ is the response of the optical detection system, calibrated with a reference deuterium lamp of known irradiance. 
\item $I_{\rm{CL}}$ is obtained by integrating the absolute CL signal in the 200-400 nm range.

\end{itemize}

Considering the whole CL procedure, the absolute LE values are considered to be measured with a 50\% uncertainty.

\section{Section II.}
 \subsection*{Thermal dissociation of excitons}

With above bandgap excitations, a part of the generated electron-hole pairs bounds under exciton quasi-particles such as $e+h\rightleftharpoons X$. This quasi-equilibrium depends on the exciton binding energy and the temperature as stated by the action mass law [9]:

\begin{equation}
	\centering
		n^{*}(T) = n_{eh}^{2}/n_{X} = (\mu k_{b}T/2\pi \hbar^{2})^{3/2}exp({-E_{b}/k_{b}T})
		\label{eq-n*}
\end{equation}

\noindent
where $n_{eh}$ and $n_{X}$ are respectively the concentrations of electron hole pairs and excitons, $\mu$ a factor including electron, hole and exciton effective masses, $E_{b}$ the exciton binding energy respectively, $k_{B}$ and $\hbar$ the Boltzmann and Planck constants. Considering the total concentration of injected carriers, free or bound as excitons, $n=n_{eh}+n_{X}$, the fraction of excitons in a crystal $f=n_{X}/n$ is:

\begin{equation}
	\centering
		f = 1 + \frac{n^{*}}{2n} - \sqrt{\Big(\frac{n^{*}}{2n}\Big)^{2} + \frac{n^{*}}{n}}
\end{equation}

$f$ depends on the temperature $T$ via $n^{*}$, but also on the injection level $n$. The equilibrium then displaces in favor of excitons at low temperatures and high injection levels. This simple model accurately describes exciton and free electron-hole pair concentrations measured with pump-probe techniques in silicon [9,10]. Implemented more recently for diamond, the thermal dissociation model successfully accounts for its luminescence properties [11].\\

As free electron-hole pairs present a much weaker radiative recombination probability compared to excitons, the temperature dependence of the diamond luminescence efficiency reflects the fraction of excitons present in the crystal. The experimental curve LE(T) is then fitted using the above expression for $f$, assuming no significant dependence of $n$ as a function of the temperature [12]. With $\mu$ = 0.076 as taken in [11], a correct fit of experimental data is obtained with $E_{b}$ = 70 meV and $n$ = 2$\times$10$^{13}$ cm$^{-3}$. The order of magnitude found for $n$ appears correct, being below the formation threshold of polyexcitons ($\sim$3$\times$10$^{13}$ cm$^{-3}$) [13] consistently with their absence of detection in the diamond CL spectrum. As mentioned in the body text, the value found for $E_{b}$ also well agrees with the standard 70-80 meV exciton binding energy determined experimentally in diamond. As a summary, the PL quenching of three orders of magnitude observed in diamond at room temperature is well fitted by the thermal dissociation of its weakly bounded excitons. For hBN we assumed $\mu$ = 0.5 and $n$ = 2$\times$10$^{14}$ cm$^{-3}$ considering a much shorter diffusion range of excitons in hBN than in diamond [12]. The curves of $f$ plotted in the Fig. 2 show the increase of $E_{b}$ required to describe the almost constant luminescence efficiency observed  over the 5-300 K temperature range in hBN. Such an analysis thus only provides a lower bound for the exciton binding energy in hBN. The present analysis indicates it is larger than 250 meV.\\

\begin{figure}[h!]
\renewcommand\thefigure{SII}
 \begin{center}
 \includegraphics[scale=0.75]{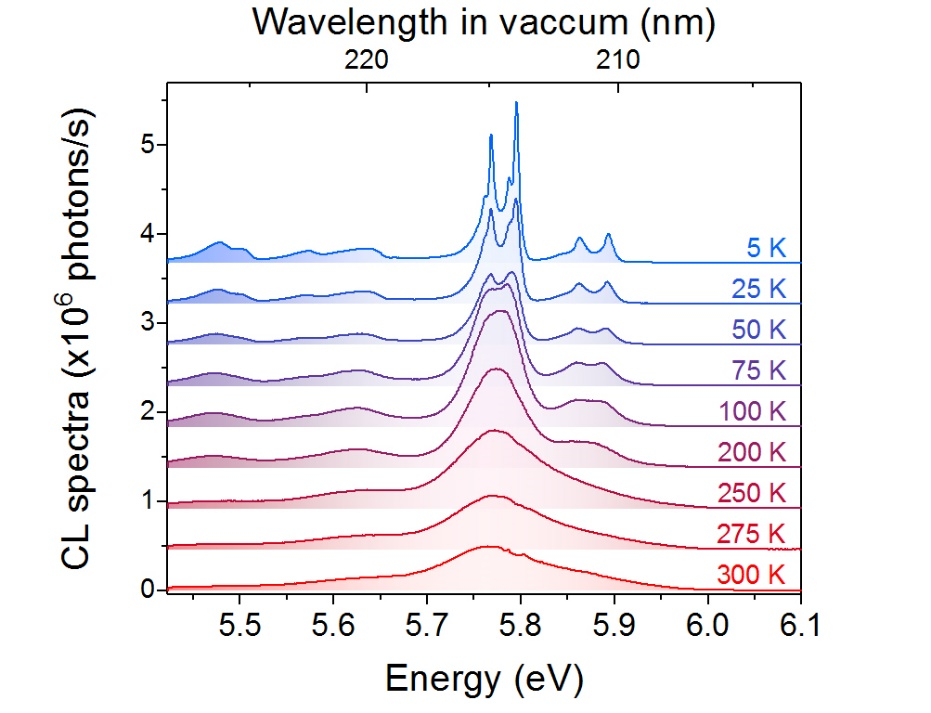}
 \caption{CL spectra of the HPHT hBN single crystal as a function of temperature from 5 K to 300 K. $V$=10 kV and $i$=0.2 nA. Spectra are shifted for clarity.}
 \label{FS2}
 \end{center}
\end{figure}

The corresponding CL spectra of hBN recorded at different temperatures are provided below (Fig.\ref{FS2}). The stability of the luminescence efficiency as a function of the temperature can be seen directly on this figure, where the integrated CL intensity appears almost constant.

\section{Section III.}
 \subsection*{Computational details}

The simulated hBN has lattice parameters $a$ = 2.5 \AA \, and $c/a$ = 2.6 [14]. The Kohn-Sham system and the GW corrections have been computed with the ABINIT simulation package (a plane-wave code [15]). Norm-conserving Troullier-Martins pseudopotentials have been used for both atomic species. DFT energies and wave functions have been obtained within the local density approximation (LDA) to the exchange-correlation potential, using a plane-wave cut-off energy of 30 Ha and sampling the Brillouin zone with a 8$\times$8$\times$4 $\Gamma$-centered k-point grid.\\

The GW quasiparticle corrections have been obtained within the G$_{0}$W$_{0}$ approach. They have been computed on all points of a 6$\times$6$\times$4 $\Gamma$-centered grid. A cut-off energy of 30 Ha defines the matrix dimension and the basis of wave functions to represent the exchange part of the self-energy. The correlation part of the self-energy has been computed including 600 bands and the same cut-off energy as for the exchange part. To model the dielectric function, the contour deformation method has been used, computing the dielectric function up to 60 eV, summing over 600 bands and with a matrix dimension of 6.8 Ha. Finally, as explained in Ref. [16], an additional shift of 0.47 eV of the empty levels (scissor operator) has been introduced to match the EELS measurements reported recently by Schuster \textit{et al.} [17]. This discrepancy has to be ascribed to lack of self-consistency in the G$_0$W$_0$ scheme. However this GW$+$Scissor scheme is preferable to a simple scissor operator because the GW self-energy is state-dependent, hence it takes into account quasiparticle effects on the dispersion of the bands and not only on their position. This is most relevant when investigating the exciton dispersion.

\begin{figure}[h!]
\renewcommand\thefigure{SIII.1}
 \begin{center}
 \includegraphics[scale=0.39]{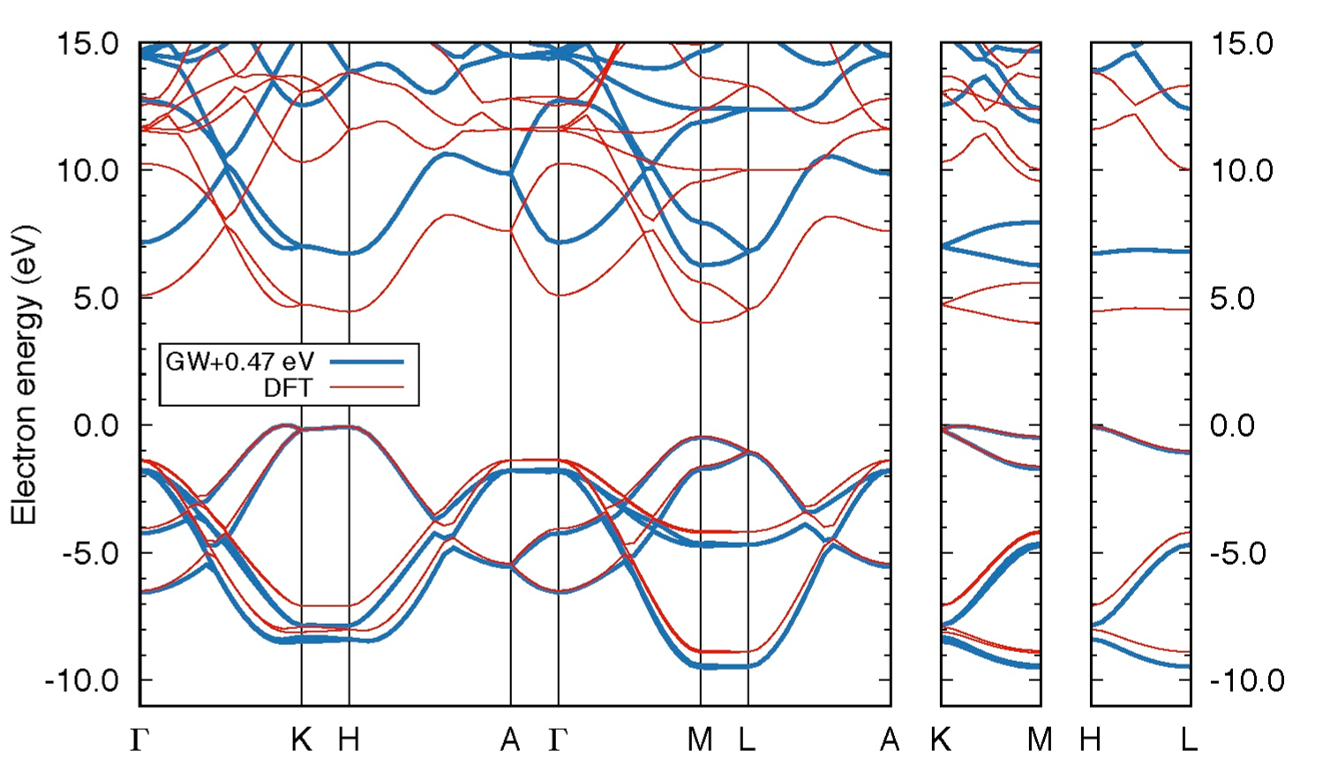}
 \caption{Electron band structure of hBN: LDA (thin red line) and G$_{0}$W$_{0}$ $+$ 0.47 eV scissor (thick blue line).}
 \label{FS3}
 \end{center}
\end{figure}

In Figure \ref{FS3}, we report the LDA and the shifted GW band structure. The latter has been obtained by making a cubic interpolation on the appropriate k-point grid of the quasiparticle corrections computed on the 6$\times$6$\times$4 grid.

The dielectric function $\epsilon(Q,\omega)$ has been calculated using the EXC [18] code on a 36$\times$36$\times$4 $\Gamma$-centered k-point grid using the shifted GW energies (properly interpolated as explained above). 
The GW-BSE has been solved in the Tamm-Dancoff approximation. Six valence bands and three conduction bands have been included in the calculation, and a cut-off energy of 360 eV for both the matrix dimension and the wave function basis has been used. The static dielectric matrix entering in the BSE kernel has been computed within the random phase approximation with local fields, including 350 bands and with cut-off energies of 120 eV and 200 eV for the matrix dimension and the wave function basis respectively. Within these parameters, the energies of the first excitons are converged within 0.01 eV.

\begin{figure}[h!]
\renewcommand\thefigure{SIII.2}
 \begin{center}
 \includegraphics[scale=0.26]{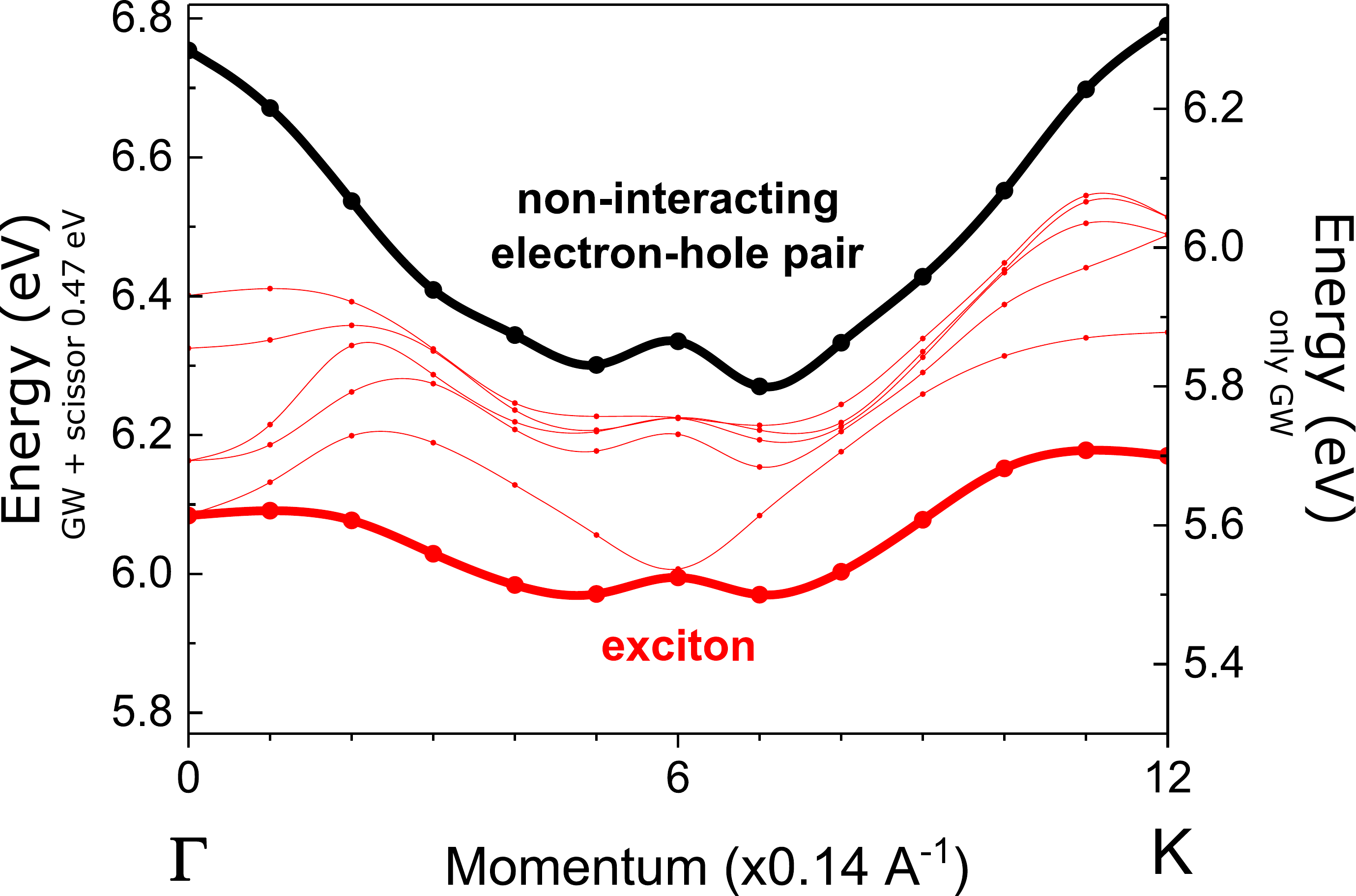}
 \caption{Dispersion of excitations for $Q$ along $\Gamma$K. Non-interacting electron-hole pair (black line) and exciton dispersion (red lines) of the first six excitons (the lowest energy one reported with a thicker line. Energy axis are GW$+$scissor (on the left) and GW only (on the right).}
 \label{FS4}
 \end{center}
\end{figure}

In Figure \ref{FS4} we report the energies of non-interacting electron-hole pair (black line) defined as:

\begin{equation}
 \centering
E_{NI}(Q) = min[E_{C}(k+Q)-E_{V}(k)]
\end{equation}

\noindent
for all $k$ in the full Brillouin zone and for $Q$ along $\Gamma$K. $E_{C}$ and $E_{V}$ are respectively the shifted GW energies of empty and conduction states. Also we report the exciton dispersion (in red), solution of the BSE (thick line = lowest laying exciton, thinner lines = higher energy excitons). The energy axis on the left includes the scissor operator of 0.47 eV, while the energy axis on the right is without the shift (only GW$+$BSE).

We point out that the lowest-energy direct exciton (at $\Gamma$) is doubly degenerate and optically dark, the closest bright exciton is about 80 meV above. The calculated values may vary depending on the approximation used. In order to give an estimate of the uncertainty we compared our data to the appendix of Ref. [16] and to other GW-BSE calculations [19]. The calculation uncertainty is estimated to be $\pm$ 50 meV for the difference between the lowest indirect and direct excitons, as well as for the indirect exciton binding energy.

